\documentstyle[prd,tighten,aps]{revtex}
\begin{document}
\draft

\twocolumn[\hsize\textwidth\columnwidth\hsize\csname
@twocolumnfalse\endcsname
\title{\bf On multiphoton absorption by molecules}

\author{Pedro F. Gonz\'alez-D\'{\i}az}
\address{Centro de F\'{\i}sica ``Miguel Catal\'an'',
Instituto de Matem\'aticas y F\'{\i}sica Fundamental,\\
Consejo Superior de Investigaciones Cient\'{\i}ficas,
Serrano 121, 28006 Madrid (SPAIN)}
\date{
arch 17, 1998}

\maketitle

\begin{abstract}
Rabi frequencies for multiphoton absorption by atoms or
molecules that can be characterized as two- or three-level
systems are obtained in the cases of single and double laser
perturbations. If the ambiguity in the origin of absolute energy
is conveniently gauged off in the three-level model, there
appears an additional non linear frequency which is originated
from the separate evolution of a relative-frequency component
induced in the whole system by the simultaneous action of the
two perturbations.

\end{abstract}

\vskip2pc]

Since multiphoton absorption and dissociation in small
polyatomic molecules were discovered two decades ago [1,2]
there has been a considerable interest [3] in the study of
the mechanism by which a single monochromatic infrared laser
radiation can be absorbed by a highly anharminic molecular mode.
This interest has led to a variety of models which range from
those suggesting a fully stochastic interpretation [4] to
models invoking rotovibrationally assisted contributions [5].
Among these models, there emerged the quite interesting
possibility that intramolecular generation of deterministic
chaos may become the main mechanism responsible for
monochromatic multiphoton
excitation of small molecules [6], with the Rabi frequency
playing the role of the control parameter. Taking the Rabi
frequency as a constant parameter could be not quite correct
an assumption as it becomes nonlinerly dependent of radiation
intensity at high intensities of the driving field [7] and
therefore no longer a genuine constant of motion.

This work aims at obtaining expressions for the Rabi frequency
in multiphoton processes using a fully quantized
procedure when the excited molecules can be modelized as 
two- or three-level systems. In accordance with previous
indications, it is obtained that such expressions generally
depend nonlinearly on field intensity, so rendering the Rabi
frequency unsuitable as constant of motion. We use two simple
models. We consider first a two-level system which coherently
absorb $M$ monochromatic photons. The results from this model
are compared with those obtained by other authors, and used
to construct the more interesting three-level model where
two independent radiation beams with different wavelengths
are considered. From this we derive the main result of this
work, that is the appearance of an additional nonlinear 
absorption frequency which can be described in terms of the
relative frequency of the two driving fields. It will be
seen that it is the existence of this additional frequency
which distinguishes a three-level system from a two-level
system with respect to multiphoton absorption.

Let us write the fully quantized Hamiltonian for the
absorption of $M$ photons with frequency $\omega$ by the
two-level system with transition frequency $\omega_{0}$
in the rotating wave approximation
\[H_{2}=H_{R}+H_{M}+H_{RM}\]
\begin{equation}
=\hbar\omega a^{\dagger}a+\frac{1}{2}\hbar\omega_{0}\sigma_{z}
+\hbar\beta\left(a^{\dagger M}\sigma_{-}-a^{M}\sigma_{+}\right),
\end{equation}
where $a$, $a^{\dagger}$, the annihilation and creation operators for
the driving field, satisfy the usual commutation relations $[a,a^{\dagger}]=1$,
and the $\sigma$'s are the $2\times 2$ Pauli matrices which, in turn,
obey the commutation relations:
\[[\sigma_{z},\sigma_{\pm}]=\pm\sigma_{\pm},\;\; [\sigma_{+},\sigma_{-}]=\sigma_{z}\]
\[\{\sigma_{+},\sigma_{-}\}=I,\]
with $I$ the unit matrix; $\beta$ is the dipole moment matrix element
for the simultaneous absorption of $M$ monochromatic photons by the
system.

The Heisenberg equations of motion read
\[\dot{\sigma}_{z}=2i\beta\left(a^{\dagger M}\sigma_{-}+a^{M}\sigma_{+}\right)\]
\begin{equation}
\dot{\sigma}_{+}=i\left(\omega_{0}\sigma_{+}-\beta a^{\dagger M}\sigma_{z}\right)
\end{equation}
\[\dot{a}=-i\left(\beta Ma^{\dagger M-1}\sigma_{-}+\omega a\right).\]

Introducing the constant of motion ($[N,H]=0$)
\begin{equation}
N=a^{\dagger}a+M\sigma_{+}\sigma_{-},
\end{equation}
we finally obtain
\[\ddot{\sigma}_{z}=-\frac{2(M\omega-\omega_{0})}{\hbar}\left[H_{2}-\hbar\omega(N-\frac{1}{2}M)\right]\]
\begin{equation}
-\left[(M\omega-\omega_{0})^{2}-2|\beta|^{2}MA_{2}(n)\right]\sigma_{z}
+\left[2|\beta|^{2}MB_{2}(n)\right]\sigma_{z}^{2},
\end{equation}
where $n=a^{\dagger}a$ and
\begin{equation}
A_{2}(n)=\frac{n!}{(n-M)!}+\frac{(n+M)!}{n!}
\end{equation}
\begin{equation}
B_{2}(n)=\sum_{\alpha=0}^{M-1}\frac{(n+\alpha)!}{(n-M+\alpha+1)!} .
\end{equation}

If we take the Rabi frequency as the coefficient of the spin operator
$\sigma_{z}$, there could be some ambiguity in such a definition using
expression (4) because the last term in this expression, which depends
on $\sigma_{z}^{2}$, could either be absorbed into the definition of
a unique Rabi frequency given by the coefficient of $\sigma_{z}$
(notice that the insertion of $\sigma_{z}=2\sigma_{+}\sigma_{-}-1$ in the
last term of (4) allows one to factorize a given coefficient for $\sigma_{z}$
depending
on $\sigma_{+}\sigma_{-}$ in that term, and then by using (3), one may
rearrange that coefficient, together with the similar one in the
second term, in such a way that the resulting overall coefficient for
$\sigma_{z}$ will not explicitely depend on operators $\sigma_{+}\sigma_{-}$
and $\sigma_{z}$, but on $N$ and $n$), or define by itself a new
nonlinear fundamental frequency if we would interpret this as
the coefficient of the spin operator squared $\sigma_{z}^{2}$. For
the case of a two-level system only the first possibility applies
because, once the physically irrelevant energy-origin gauge is
substracted off, we are left with just one single periodic
evolution for the two eigenstates in the semiclassical treatment;
i.e.: for two-level systems, one can always have an one-body
representation of the pure molecular Hamiltonian and this should
associate with a single Rabi frequency given by
\[\Omega_{R}^{2}=(M\omega-\omega_{0})^{2}\]
\begin{equation}
+2|\beta|^{2}\left[MA_{2}(n)-B_{2}(n)\left(M-2N+2n\right)\right].
\end{equation}
Thus, $\Omega_{R}$ is no longer a constant of motion and
depends explicitly on $n$ nonlinearly for $M>1$. Similar
expressions have been obtained by Sukumar and Buck [8] and
Kochetov [9]. We note that expression (7) reduces to the
known Knight-Milonni equation [10] for the case of linear absorption
$M=1$.

The existence of only a single linear Rabi frequency, $\Omega_{R}$,
is no longer a characteristic of multiphoton absorption by a
N-level system when N$>2$. In dealing with the molecular
Hamiltonian as a level system in terms of the absolute energies of
the levels, one always has an ambiguity in the choice of the
origin for such energies. This ambiguity is eliminated by re-expressing
the Hamiltonian in terms of the level transition energies rather than
absolute energies. In the case of a three-level system, N=3, this
can be done by writing $H_{M}$ as (Fig. 1)
\begin{equation}
H_{M}=\frac{1}{2}\left\{(E_{1}+E_{3})I+(E_{3}-E_{1})S
+\left[2E_{2}-(E_{1}+E_{3})\right]S_{5}\right\},
\end{equation}
where
\begin{equation}
S=
\left(
\begin{array}{ccr}
 1&0&0\\
      0&0&0\\
      0&0&-1
\end{array}
\right)
\end{equation}
and
\begin{equation}
S_{5}=
\left(
\begin{array}{ccc}
 0&0&0\\
      0&1&0\\
      0&0&0
      \end{array}
      \right)
      \end{equation}

Then, since $S_{5}=I-S^{2}$, (8) transforms into:
\begin{equation}
H_{M}=E_{2}I+\frac{1}{2}\hbar(\omega_{0}+\omega_{1})S-
\frac{1}{2}\hbar(\omega_{0}-\omega-{1})S^{2}.
\end{equation}
The matrix $I$ can be dropped off from (11) by a convenient shift of
the energy origin.Taking this origin at the energy of the intermediate
level $E_{2}$, we have
\begin{equation}
H_{M}=\frac{1}{2}\hbar(\omega_{0}+\omega_{1})S-
\frac{1}{2}\hbar(\omega_{0}-\omega-{1})S^{2}.
\end{equation}
Thus, much as it is usually done in the two-body problem of mechanics,
we have reduced the three level problem to the separate evolution of
the center of energies and the relative energy. Note that for the
harmonic-oscillator case only the term depending on $(\omega_{0}+\omega_{1})$
survives. When $\omega_{0}\neq\omega_{1}$ there would then be an additional
radiative mechanism to be superimposed to the usual one. Hence, two
independent contributing frequencies should be expected.

The fully quantized Hamiltonian for the three-level molecular
system depicted in Fig. 1, for the case that we irradiate with
two beams of frequencies $\omega_{L1}$ and $\omega_{L2}$ becomes
\[H_{3}=\hbar\omega_{L1}a_{1}^{\dagger}a_{1}+\hbar\omega_{L2}a_{2}^{\dagger}a_{2}\]
\[+\frac{1}{2}\hbar\left[(\omega_{0}+\omega_{1})S-(\omega_{0}-\omega_{1})S^{2}\right.\]
\begin{equation}
\left.+2\beta\left(a_{1}^{\dagger M}a_{2}^{\dagger(N-M)}S_{8}S_{4}-a_{1}^{M}a_{2}^{(N-m)}S_{2}S_{6}\right)\right],
\end{equation}
where $\beta$ is now the dipole moment matrix element for the total
absorption of the $N$ photons, $M$ with energy $\hbar\omega_{L1}$
and $N-M$ with energy $\hbar\omega_{L2}$. The operators $S_{i}$
are the generators of the unitary group $U(3)$ and obey
\begin{equation}
[S_{i},S_{j}]=C^{k}_{ij}S_{k},
\end{equation}
in which $i,j,k=1,2,...,9$, and $C^{k}_{ij}$ are the structure constants.

Introducing the constants of motion ($[N_{1},H_{3}]=[n_{2},H_{3}]=0$)
\[N_{1}=a_{1}^{\dagger}a_{1}+\frac{M}{2}S\]
\begin{equation}
N_{2}=a_{2}^{\dagger}a_{2}+\frac{n-M}{2}S,
\end{equation}
from the Heisenberg equations, we obtain finally
\begin{equation}
\ddot{S}=\frac{\epsilon}{\hbar}(H_{3}-\hbar\omega_{L1}N_{1}-\hbar\omega_{L2}N_{2})
\Omega_{Rc}^{2}S+\Omega_{Rr}^{2}S^{2},
\end{equation}
where
\[\epsilon=2\left[\omega_{0}+\omega_{1}-M\omega_{L1}-(N-M)\omega_{L2}\right]\]
\[\Omega_{Rc}^{2}=-\frac{\epsilon}{2}(\omega_{0}+\omega_{1}-M\omega_{L1})\]
\begin{equation}
+\frac{\epsilon}{2}(N-M)\omega_{L2}-|\beta|^{2}A_{3}(n_{1},n_{2})
\end{equation}
\begin{equation}
\Omega_{Rr}^{2}=-\frac{\epsilon}{2}(\omega_{1}-\omega_{0})-2|\beta|^{2}B_{3}(n_{1},n_{2})
\end{equation}
with
\[A_{3}(n_{1},n_{2})=\frac{n_{1}!n_{2}!}{(n_{1}-M)!(n_{2}-N+M)!}\]
\[+\frac{(n_{1}+M)!(n_{2}+N-M)!}{n_{1}!n_{2}!},\]
\[B_{3}(n_{1},n_{2})=M\frac{n_{2}!}{(n_{2}-N+M)!}\sum_{\alpha=0}^{M-1}\frac{(n_{1}+\alpha)!}{(n_{1}-M+\alpha+1)!}\]
\[+(N-M)\frac{n_{1}!}{(n_{1}-M)!}\sum_{\gamma=0}^{N-M-1}\frac{(n_{2}+\gamma)!}{(n_{2}-N+M+\gamma+1)!},\]
and $n_{i}=a_{i}^{\dagger}a_{i}$.

The roots of the coefficients for the matrices $S$ and $S^{2}$ can be
consistently interpreted, respectively, as the Rabi frequency for
the center of energies $\Omega_{Rc}$, and the Rabi frequency for
the relative energy $\Omega_{Rr}$. We see that (15) and (16) are
again radiation-intensity dependent.

For $N=M=1$ in the limit $\omega_{1}=\omega_{L1}\rightarrow 0$,
$S\rightarrow\sigma_{z}$ and $\beta\rightarrow f$, so that the
whole three-level system reduces to an {\it effective} two-level
system, and $\Omega_{Rr}$ becomes a constant to be absorbed, one
part, $-2|f|^{2}$, into the conventional Rabi frequency, and the
other part, $\omega_{0}(\omega_{0}-\omega_{L1})$, into the
counter-part of the first term in the r.h.s. of (14) which contains
the constants of motion. This seems to suggest that $\Omega_{Rr}$
ultimately originates from a nonlinear contribution to
radiation-matter interaction through which both radiation beams
and the matter system would all cooperate. In fact, the only
contributions that can survive the
rotating wave approximation with respect to
$\omega_{0}-\omega_{1}$ in the semiclassical treatment (where
we consider a quantized matter system being perturbed by a
classical radiation field [11]) are those arising
from nonlinear terms which simultaneously contain both radiation
field perturbations. Thus, one should regard the presence of
the separately-operating frequencies $\Omega_{Rc}$ and $\Omega_{Rr}$
as a consequence from the full-quantized picture.
In the semiclassical treatment, possible somewhat analogous
effects arising from the novel non-linear Rabi frequency could
still be found, though they would not be expected to be
visualizable as being driven from two idependent Rabi
frequencies, but rather from a more complicate single Rabi
parameter.

The incidence that one could expect from the above predictions
on multiphoton absorption of polyatomic molecules is twofold.
On one hand, the new Rabi frequency would provide us with a novel
mechanism to help overcoming the molecular anharmonic barrier
operating in the few low-lying levels; on the other hand, such
a Rabi frequency would play the role of still another control parameter
in models where dichromatic multiphoton excitation is assumed to
be driven by intermolecular generation of deterministic chaos,
allowing for a more effective coupling of the active excited
levels to a harmonic reservoir, and hence leading to an increase
of the multiphoton absorption cross section.

In summary, starting with the consideration of multiphoton
absorption by a two-level system, the present report contains
the derivation of two fundamental frequencies in order to
characterize a two-frequency laser induced multiabsorption
by a three-level system. Such frequencies are radiation-intensity
dependent and correspond to the center of energy and relative
energy, respectively.

\acknowledgements

\noindent For helpful comments and help in the calculations,
the author thanks C. Sigenza
and M. Santos. This
research was supported by DGICYT under research projects N§
PB94-0107 and N§ PB93-0139.


\pagebreak

\begin{center}
{\bf Legend for Figures}
\end{center}

\vspace{.5cm}

\noindent $\bullet$ Fig. 1. Relevant transitions in a three-level system.


\begin{references}

\bibitem {1} N.R. Isenor, V. Merchant, R.S. Hallsworth and M.C. Richardson,
{\it Can. J. Phys.} 51, 1281 (1973).
\bibitem {2} R.V. Ambartzumian, N.V. Chekalin, V.S. Doljikov, V.S. Letokhov
and E.A. Ryabov, {\it Chem. Phys. Lett.} 25, 515 (1974).
\bibitem {3} See for example {\it Multiple-Photon Excitation and Dissociation
of Polyatomic Molecules} (Springer, Berlin, 1986); S.S. Mitra and
S.S. Bhatlacharyya, {\it J. Phys.} B25, 2535 (1992)
\bibitem {4} S. Mukamel, {\it J. Chem. Phys.} 70, 5834 (1979).
\bibitem {5} C.D. Cantrell and H.W. Galbraith, {\it Opt. Commun.} 18, 513 (1976);
21, 374 (1977); D.M. Larsen and N. Bloembergen, {\it Opt. Commun.} 17, 250 (1976).
\bibitem {6} J.R. Ackerhalt, H.W. Galbraith and P.W. Malonni, {\it Phys. Rev. Lett.}
51, 1259 (1983).
\bibitem {7} X-s Li and N-y Bei, {\it Phys. Lett.} 101A, 169 (1984).
\bibitem {8} C.V. Sukumar and B. Buck, {\it J. Phys.} A17, 885 (1984).
\bibitem {9} E.A. Kochetov, {\it J. Phys.} A20, 2442 (1987).
\bibitem {10} P.L. Knight and P.W. Milonni, {\it Phys. Rep.} 66, 21 (1980).
\bibitem {11} M. Schubert and B. Wilhelmi, {\it Nonlinear Optics
and Quantum Electronics} (John Wiley and Sons, New York, 1986), p. 458.


\end{references}
\end{document}